\documentclass[prm, twocolumn,amsmath,amssymb]{revtex4}

\usepackage{graphicx}
\usepackage{bm}
\usepackage{gensymb}
\usepackage{multirow}
\usepackage{xcolor}

\DeclareGraphicsExtensions{.eps, .jpg,.pdf}

\begin{document}

\title{Influence of cobalt substitution on the magnetism of NiBr$_2$\footnote{
Notice: This manuscript has been authored by UT-Battelle, LLC under Contract No. DE-AC05-00OR22725 with the U.S. Department of Energy. The United States Government retains and the publisher, by accepting the article for publication, acknowledges that the United States Government retains a non-exclusive, paid-up, irrevocable, world-wide license to publish or reproduce the published form of this manuscript, or allow others to do so, for United States Government purposes. The Department of Energy will provide public access to these results of federally sponsored research in accordance with the DOE Public Access Plan (http://energy.gov/downloads/doe-public-access-plan).}}

\author{Binod K. Rai$^1$}
\email[]{rairk@ornl.gov}
\author{Andrew D. Christianson$^1$}
\author{David Mandrus$^{1,2,3}$}
\author{Andrew F. May$^1$}

\affiliation{$^1$Materials Science and Technology Division, Oak Ridge National Laboratory, Oak Ridge, TN 37831, USA}
\affiliation{$^2$Department of Physics and Astronomy, University of Tennessee, Knoxville, Tennessee 37996, USA}
\affiliation{$^3$Department of Materials Science and Engineering, University of Tennessee, Knoxville, Tennessee 37996, USA}
\date{\today}

\begin{abstract}
Co substituted Ni$_{1-x}$Co$_x$Br$_2$ (0 $\leq~x~\leq$ 1) single crystals were synthesized using vapor transport. The physical properties of the crystals were characterized by x-ray powder diffraction, magnetization, and specific heat measurements. Room temperature x-ray powder diffraction data indicate a change from the CdCl$_2$ structure type to the CdI$_2$ structure type occurs within 0.56 $<~x~<$ 0.76. NiBr$_2$ has a commensurate antiferromagnetic phase below $T_{\rm N} \approx$ 46 K and an incommensurate magnetic ground state below $T_{\rm IC} \approx$ 20 K. Both magnetic transitions are affected by cobalt substitution, and the incommensurate phase transition is present up to at least $x$ = 0.56. The evolution of magnetism has been studied as a function of cobalt content and is summarized in the temperature-composition phase diagram.

\end{abstract}

\maketitle

\section{Introduction}
Binary transition metal dihalides $MX_2$ ($M$ = transition metal and $X$ = halogen) are of great interest to the low-dimensional magnetism community because of their novel physics and potential applications\cite{Mushtaq2017,McGuire2017, Chen2017}. $MX_2$ compounds contain triangular nets of transition metal ions, and a helimagnetic ground state is observed in the few dihalides where long-range interactions are important\cite{Adam1980, Kuindersma1981, Iio1990}. The dihalides that host the helimagnetic structure have received attention in recent years due to the discovery of the coupling between helimagnetic order and ferroelectric polarization\cite{Kurumaji2011, Tokunaga2011, Wu2012, Kurumaji2013}. NiBr$_2$ has recently received renewed interest as a multiferroic\cite{Tokunaga2011} and for its potential to host a magnetic vortex state upon chemical substitution\cite{Lin2016} or multiple-$q$ states under applied magnetic field\cite{Okubo2012,Babu2018}, all of which are associated with its helimagnetic  ground state.

\begin{figure}[ht!]
	\includegraphics[clip,width=\columnwidth]{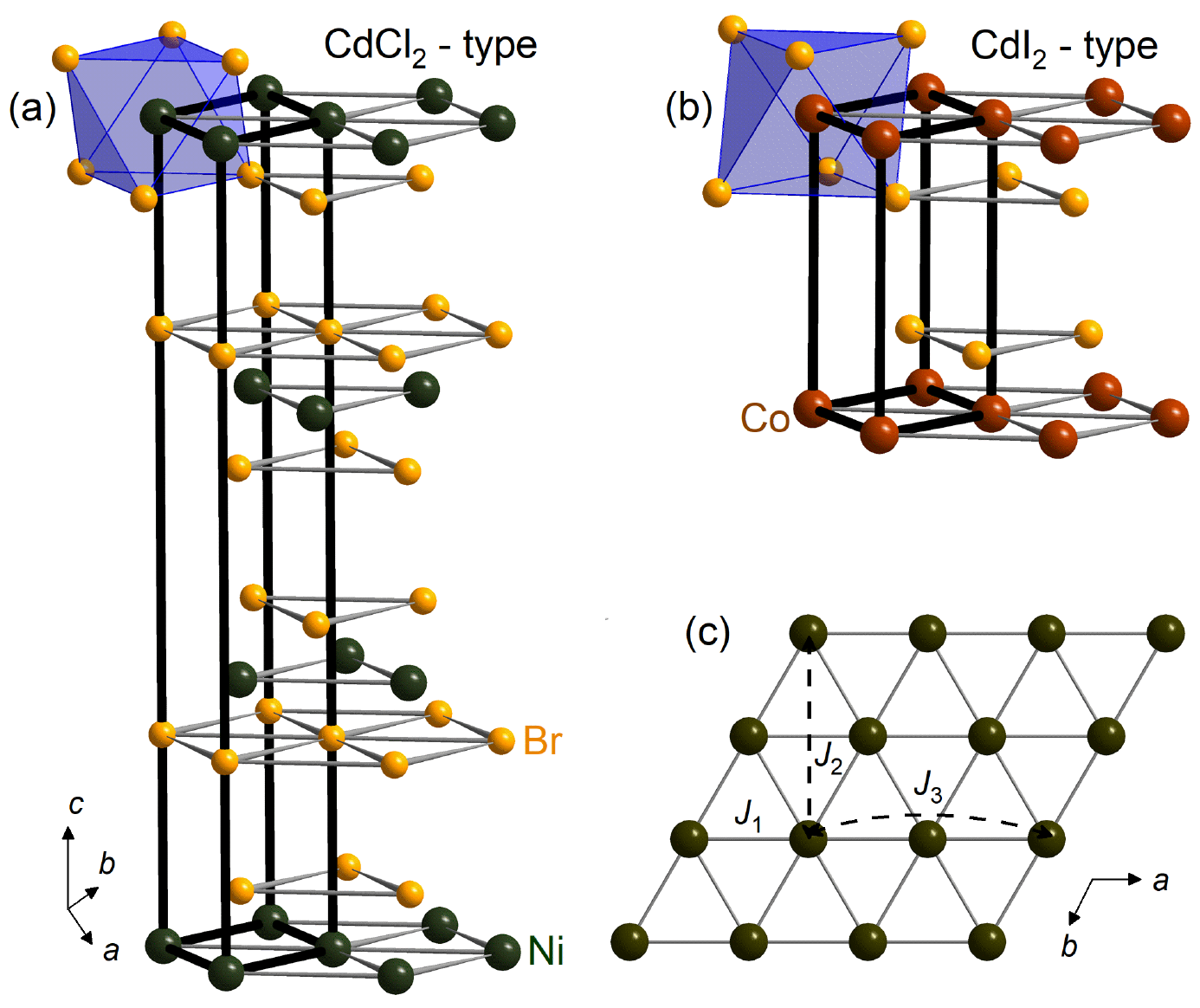}
\caption{\label{Fig} (a) NiBr$_2$ crystal structure with the CdCl$_2$ structure type ($R\bar{3}m$). (b) CoBr$_2$ crystal structure with the CdI$_2$ structure type ($P\bar{3}m1$). In (a,b), a single unit cell is outlined and lines connect atoms to aid in viewing. (c) Plan view of a single layer of transition metal with exchange constants $J_{1,2,3}$ defined.}
\end{figure}

NiBr$_2$ is a magnetic insulator that has a commensurate magnetic structure below an antiferromagnetic transition temperature of $T_{\rm N} \approx$ 48 K\cite{Day1981Moore, Moore1985, Tokunaga2011}. The magnetic structure changes from a commensurate one to an incommensurate one below $T_{\rm IC} \approx$ 20 K, and this magnetic ground state is a helical spin structure\cite{Day1981Moore,Adam1980}. In both magnetic phases, the moments are oriented within the basal plane of the rhombohedral structure (see Fig.\ref{Fig}). The incommensurate structure is believed to result from a delicate balance between the first-neighbor ferromagnetic exchange constant $J_1$ and the longer-range antiferromagnetic exchange constants $J_2$ and $J_3$ (see Fig. \ref{Fig} (c))\cite{Rastelli1979,Odisho1989, Leonov2015}. The competition between these exchange constants is affected by external control parameters including magnetic field, pressure, or chemical substitution\cite{Katsumata1983, Yoshiyama1984, Yoshiyama1986,Day1981Moore, Day1981,Adam1981,Regnault1982,Moore1985}, which destabilize the incommensurate spin structure. An applied magnetic field suppresses $T_{\rm IC}$, and the incommensurate phase changes to the commensurate one around 2.7 T at $T$ = 2 K for $H\perp c$\cite{Katsumata1983}. The helimagnetic structure of NiBr$_2$ is also affected by hydrostatic pressure and it disappears with applied pressure above 10.6 kbar, while the antiferromagnetic phase remains unchanged\cite{Day1981, Adam1981}. 

Chemical substitution studies of NiBr$_2$ have been reported and a complete Ni-site substitution by other transition metals (Ni$_{1-x}TM_x$Br$_2$, $TM$ = Mn, Fe, Zn) was hindered by solubility limits\cite{Moore1981,Day1981,Moore1985}. Fe-doping suppresses $T_{\rm N}$, and Fe content greater than $x \geq$ 0.112 is reported to eliminate the incommensurate phase and induce an easy-axis, collinear antiferromagnetic structure\cite{Moore1985}. The highest reported Mn content is $ x \approx$ 0.03, which slightly suppresses $T_{\rm IC}$ without changing $T_{\rm N}$\cite{Moore1985}. In Ni$_{1-x}$Zn$_x$Br$_2$, the highest reported Zn content $x$ = 0.08 slightly suppresses $T_{\rm N}$ and $T_{\rm IC}$. Interestingly, the propagation vector of the incommensurate structure becomes disordered upon Zn substitution\cite{Moore1981,Day1981}, and it has been recently proposed that Ni$_{0.92}$Zn$_{0.08}$Br$_2$ could contain an impurity-driven vortex lattice phase\cite{Lin2016}. These results highlight the sensitivity of the incommensurate spin structure of NiBr$_2$ to chemical perturbation.

To further explore the magnetism of NiBr$_2$, we have investigated the impact of cobalt substitution in Ni$_{1-x}$Co$_x$Br$_2$. CdCl$_2$ and CdI$_2$ are the prototypes for the two dominant structure types in the $MX_2$ family, shown in Fig. \ref{Fig}. In both structure types, triangular nets of transition metal ions are separated by two planes of halide ions that are weakly bonded by van der Waals forces\cite{McGuire2017}. The main difference between these structure types is the layer stacking, which is reflected in the lattice centering. NiBr$_2$ crystallizes in the rhombohedral CdCl$_2$ structure type with space group $R\bar{3}m$\cite{Adam1980} and has $ABC$ layer stacking yielding three NiBr$_2$ layers per unit cell. CoBr$_2$ crystallizes in the CdI$_2$ structure type with space group $P\bar{3}m1$, and it has one CoBr$_2$ layer per unit cell\cite{Wilkinson1959,Katsumata1984,McGuire2017}. Thus, the $c$-axis lattice parameter of NiBr$_2$ is approximately three times that of CoBr$_2$. CoBr$_2$ has a commensurate magnetic structure below $T_{\rm N} \approx$ 19 K and does not host a helimagnetic ground state\cite{Wilkinson1959,Katsumata1984}.

In this study, single crystals of Ni$_{1-x}$Co$_x$Br$_2$ (0 $\leq~x~\leq$ 1) have been synthesized and characterized using magnetization and specific heat measurements. A composition-induced transition from the CdCl$_2$ structure type of NiBr$_2$ to the CdI$_2$ structure type of CoBr$_2$ has been observed within 0.56 $<~x~<$ 0.76 by room temperature powder x-ray diffraction data. A magnetic phase diagram has been established based on magnetization and specific heat measurements. For samples with the CdCl$_2$ structure type, $T_{\rm N}$ is continuously suppressed with increasing cobalt concentration. The stability of the incommensurate phase seems to be enhanced for $x \leq$ 0.26, and signatures of this phase are detected up to $x$ = 0.56.  

\section{Experimental Methods}

\begin{figure}[htbp]
	\includegraphics[clip,width=0.9\columnwidth]{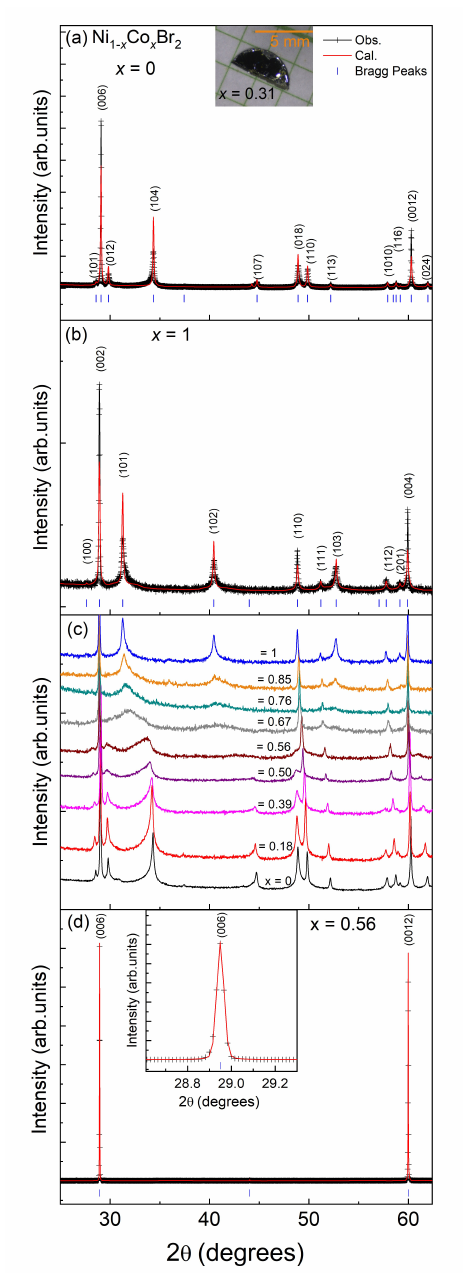}
\caption{\label{Fig1a} Room temperature x-ray powder diffraction pattern (black line + symbol) with Le Bail fit (red line) of (a) NiBr$_2$ and (b) CoBr$_2$. Inset of (a): crystal picture for $x$ = 0.31. (c) X-ray diffraction data of Ni$_{1-x}$Co$_x$Br$_2$. (d) A representative diffraction pattern from a facet for $x$ = 0.56. Inset of (d): zoom in view of 006 reflection.   }
\end{figure}

Single crystals of Ni$_{1-x}$Co$_x$Br$_2$ (0 $\leq~x~\leq$ 1) were grown from anhydrous NiBr$_2$ (99.99\% purity) and CoBr$_2$ (99.99\% purity) using vapor transport. NiBr$_2$ and CoBr$_2$ were placed in a silica ampoule inside a helium-filled glovebox and sealed under vacuum without exposure to air. The ampoules were heated in a horizontal tube furnace at temperatures ranging from 900 $^{\circ}$C (NiBr$_2$) to 620 $^{\circ}$C (CoBr$_2$) for several days, and cooled to room temperature within 1 day. The growths resulted in crystals at the cold end of the ampoule, with crystalline facets that were the cleavage $ab$-plane. Despite optimizing the growth conditions for each composition, the size of the single crystals decreased with increasing $x$; a representative crystal ($x \approx$ 0.31) is shown in the inset of Fig. \ref{Fig1a} (a). The Ni$_{1-x}$Co$_x$Br$_2$ crystals are air sensitive and the sensitivity to air increases as Co content increases. We emphasize that care must be taken to avoid air exposure, especially for large $x$, or else anomalous physical properties may be observed below 10 K.

Room temperature powder x-ray diffraction data were collected in a PANalytical X'Pert Pro MPD diffractometer (Cu K$\alpha_1$ radiation). The crystals were mixed with glass powder, and the mixtures were ground into powders using a mortar and pestle inside a helium glovebox. The fine powders were sealed in an air-free sample holder. X-ray diffraction data were also collected from facets of freshly-cleaved single crystals, and these data contain the 00l reflections. The x-ray diffraction data were analyzed with the program FullProf\cite{Rodriguez-Carvajal} using the Le Bail technique. Energy dispersive x-ray spectroscopy (EDS) was employed to determine the $x$ values from relative Ni and Co concentrations assuming full transition metal occupancy; a Hitachi TM3000 with Bruker EDS detector was utilized. EDS was performed on both sides of at least two crystals for each $x$, and the EDS values were obtained by averaging the results of at least 20 spots per measurement with error bars being the standard deviation. The $x$ values provided in this work are the experimental ones obtained from EDS and are reported in the Table~\ref{Table_EDS}. The EDS measurements also provide important guidance on the growth optimization, as samples from growths that were not properly optimized exhibited a large variation in $x$. Anisotropic magnetization measurements were performed in a Quantum Design Magnetic Property Measurement System. To identify intrinsic magnetic behavior for the most air-sensitive samples, including CoBr$_2$, the crystals were sealed inside a quartz tube containing helium exchange gas. Specific heat data were collected in a Quantum Design Physical Property Measurement System.

\begin{figure}[h!]
	\includegraphics[clip,width=0.9\columnwidth]{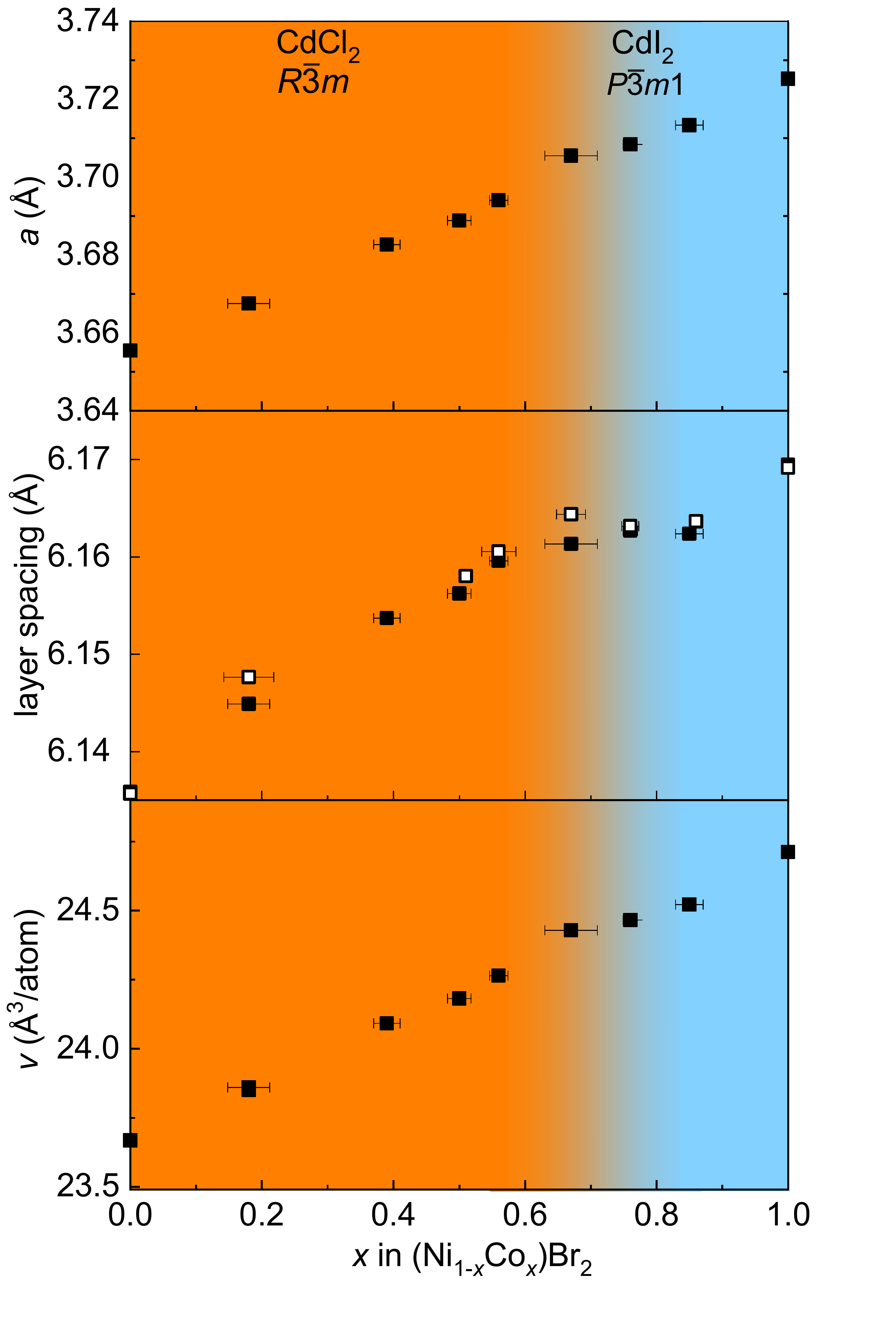}
\caption{\label{Fig1b} (a) Lattice parameter $a$, (b) layer spacing, and (c) the average volume per atom as a function of $x$ (obtained from EDS) in Ni$_{1-x}$Co$_x$Br$_2$ at 300 K. The vertical error bars are smaller than the data points. Note: closed symbols from powder diffraction and open symbols for diffraction data collected from the facets of single crystals. The area where the colors merge represent a region where stacking disorder hinders the ability to define a macroscopic symmetry.}
\end{figure}

\section{Results and Discussion}
\subsection{Structure}
Room temperature x-ray powder diffraction data are shown in Fig. \ref{Fig1a}. The NiBr$_2$ (see Fig. \ref{Fig1a} (a)) and CoBr$_2$ (see Fig. \ref{Fig1a} (b)) powder diffraction patterns are well described by the CdCl$_2$ structure type and the CdI$_2$ structure type, respectively, using Le Bail fitting. Figure \ref{Fig1a} (c) shows the powder diffraction patterns of Ni$_{1-x}$Co$_x$Br$_2$. The diffraction data show the change in structure type at room temperature occurs within 0.56 $<~x~<$ 0.76. This transition is best viewed by tracking the 104 and 102 reflections of NiBr$_2$ and CoBr$_2$, respectively.

The asymmetry and broadness of the Bragg reflections increase as $x$ approaches the change in the structure type. Indeed, the weakest h0l reflections are fully suppressed near the change in the structure type. The apparent asymmetric and broad Bragg reflections in such layered materials may result from a combination of stacking faults or disorder or strain \cite{Hendricks1942,Ramesh2003}. In our study, the 00l and hhl peaks obtained from diffraction data are not significantly broadened, even for $x$ = 0.67. The h0l reflections are broadened and this suggests stacking disorder contributes most to the broadening\cite{Ramesh2003}. The compositions around the change in the structure type likely have a significant amount of disorder associated with layer stacking, and this makes defining a precise composition where the structure type changes inappropriate because layer stacking is what differentiates the structure types. For the $x$ = 0.67 sample where a large degree of stacking disorder and composition variation are evident, we obtained lattice parameters using the CdI$_2$ model because it provided the best fitting of the diffraction data. However, near this composition it is probably not reasonable to ascribe a particular structure model.





 \begin{figure}[t!]
	\includegraphics[clip,width=0.9\columnwidth]{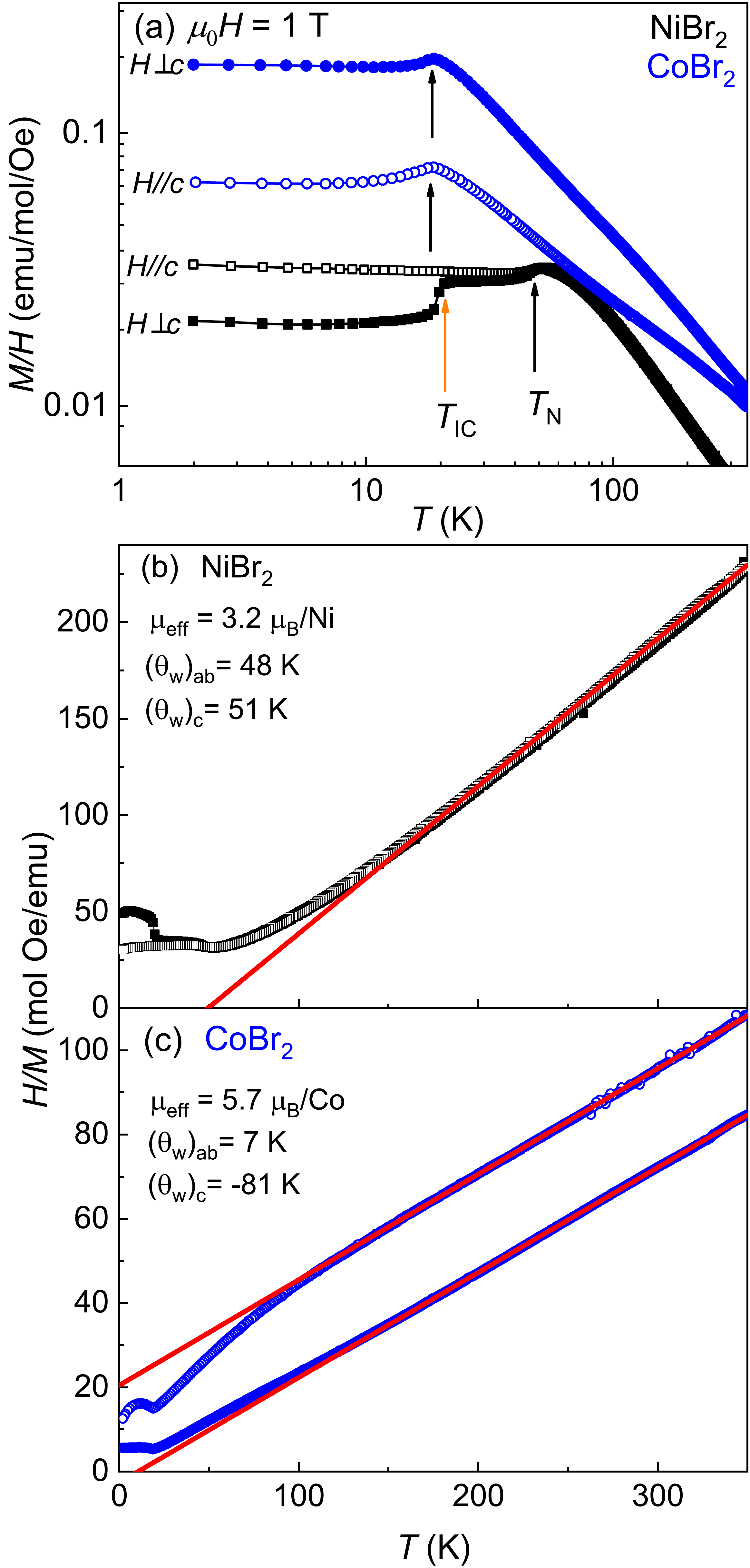}
\caption{\label{Fig2} (a) Temperature-dependent magnetization $M/H$ collected upon cooling in an applied magnetic field $\mu_0 H$ = 1 T for NiBr$_2$ (black) and CoBr$_2$ (blue). Temperature-dependent inverse magnetic susceptibility $H/M$ for (b) NiBr$_2$ and (c) CoBr$_2$, with the Curie-Weiss fit between 150 K $\leq~T~\leq$ 350 K shown by red line.}
\end{figure}

\begin{table}[h]
	\begin{tabular}{|c|c|c|c|c|c|c|c|c|c|c|c|c|}
		\hline
		Nominal         & 0.10   & 0.20 & 0.25 & 0.30 & 0.40 & 0.5 & 0.55 & 0.70 & 0.75 & 0.80 \\ \hline
		EDS$~x$         & 0.11   & 0.18 & 0.25 & 0.31 & 0.39 & 0.5 & 0.56 & 0.67 & 0.76 & 0.85 \\ \hline
		STD ($\%$) & 1.0   & 3.8 & 1.5 & 2.6 & 2.0 & 1.8 & 1.4 & 4.0 & 1.1 & 2.1\\ \hline
	\end{tabular}
	\centering
	\caption{The nominal growth composition, the value obtained by EDS, and the corresponding standard deviations (STD) that provide the relative error for the EDS measurements.}
	\label{Table_EDS}
\end{table}

The change in the lattice parameters as a function of $x$ is shown in Fig. \ref{Fig1b}. The $a$-axis lattice parameter increases continually from $x$ = 0 to 1 as shown in Fig. \ref{Fig1b} (a). The layer spacing and the average volume per atom increase with increasing $x$ similar to the change in the $a$-axis lattice parameter (see Fig. \ref{Fig1b} (b,c). The layer spacing is equal to $c$ for samples with the CdI$_2$ structure type and is equal to $c$/3 for those with the CdCl$_2$ structrure type. Le Bail fitting (2$\theta$ = 10$^\circ$ - 100$^\circ$) was performed to obtain the lattice parameters of Ni$_{1-x}$Co$_x$Br$_2$, and the reported lattice parameters for $x$ = 0.56 and 0.67 were calculated excluding (h0l) Bragg reflections from the fitting. 

X-ray diffraction data collected from the facets of single crystals are characterized by sharp 00l reflections for all samples examined, including $x$ = 0.56 (shown in Fig. \ref{Fig1a} (d)). This suggests the crystals do not have significant macroscopic chemical inhomogeneity or immiscibility. The expansion of the layer spacing obtained by fitting these data is in agreement with that obtained from the powder diffraction data.  We note, however, that the $x$ = 0.67 sample has much higher standard deviations in the EDS measurements and the symmetry is not well-defined. Thus, it is possible that near $x$ = 0.67 some immiscibility exists that is beyond our resolution.

\subsection{Magnetic Properties}

The temperature-dependent anisotropic magnetization $M$ of NiBr$_2$ and CoBr$_2$ was measured in an applied magnetic field $\mu_0 H$ = 1 T, and the results are shown in Fig. \ref{Fig2}. A broad cusp in the temperature-dependent magnetization $M/H$ is observed upon cooling, which corresponds to a transition from a paramagnetic phase to a commensurate antiferromagnetic phase. The transition temperatures $T_{\rm N}$ = 46(1) K for NiBr$_2$ and 18(1) K for CoBr$_2$ are determined from peaks in d($MT$)/d$T$\cite{Fisher1962}. In addition to $T_{\rm N}$ in NiBr$_2$, when $H\perp c$, a small decrease in $M/H$ below $T \approx$ 20 K is observed and corresponds to the onset of the helimagnetic structure. The transition temperature $T_{\rm IC}$ = 20(1) K is also determined from a peak in d($MT$)/d$T$. $T_{\rm N}$ and $T_{\rm IC}$ determined from our study are in agreement with previous reports\cite{Wilkinson1959,Day1981}. $T_{\rm IC}$ is not observed when $H\parallel c$ for NiBr$_2$, and it is absent in CoBr$_2$. We note that $T_{\rm IC}$ of NiBr$_2$ is suppressed with applied magnetic field $H\perp c$\cite{Tokunaga2011}.

Neutron diffraction studies on NiBr$_2$ and CoBr$_2$ reported that the moments align ferromagnetically within a layer and stack antiferromagnetically along the $c$ axis, and furthermore the studies suggest that the ordered moments orient in the $ab$ plane\cite{Wilkinson1959, Day1976, Day1988}. Despite this similarity from the perspective of neutron diffraction, the measured anisotropy of the induced magnetization is opposite between NiBr$_2$ and CoBr$_2$ (see Fig. \ref{Fig2}(a)). The temperature-dependent $M/H$ of both compounds shows Curie-Weiss behavior at high temperature. The effective moments and Weiss temperature $\theta_W$ are extracted from a linear fit (red line) of the inverse magnetic susceptibility $H/M$ at 150 K $\leq~T~\leq$ 350 K, as shown in Fig.\ref{Fig2}(b,c). For NiBr$_2$, the experimental effective moment is 3.25 $\mu_B$/Ni, which is larger than the calculated spin-only ($S$ = 1) effective moment of 2.83 $\mu_B$/Ni$^{+2}$. The experimental effective moment suggests some orbital contribution to the moment, which is consistent with the reported $g~\approx$ 2.2 of NiBr$_2$\cite{Tuchendler1985}. For NiBr$_2$, we obtained $\theta_W$ of 48 K and 51 K for $H\perp c$ and $H\parallel c$, respectively, indicating ferromagnetic correlations in the paramagnetic state that are consistent with a previous report\cite{Regnault1982}. For CoBr$_2$, the experimental effective moment of 5.53  $\mu_B$/Co is larger than the calculated spin-only value (3.87 $\mu_B$/Co$^{+2}$), and a similar effective moment of 5.29 $\mu_B$/Co has been reported in isostructural CoCl$_2$\cite{Starr1940}. The $\theta_W$ are 7 K and -81 K for $H\perp c$ and $H\parallel c$, respectively, and this behavior may be driven by single ion anisotropy and/or anisotropic exchange interactions\cite{Johnston2017}.

\begin{figure}[htbp]
	\includegraphics[clip,width=0.9\columnwidth]{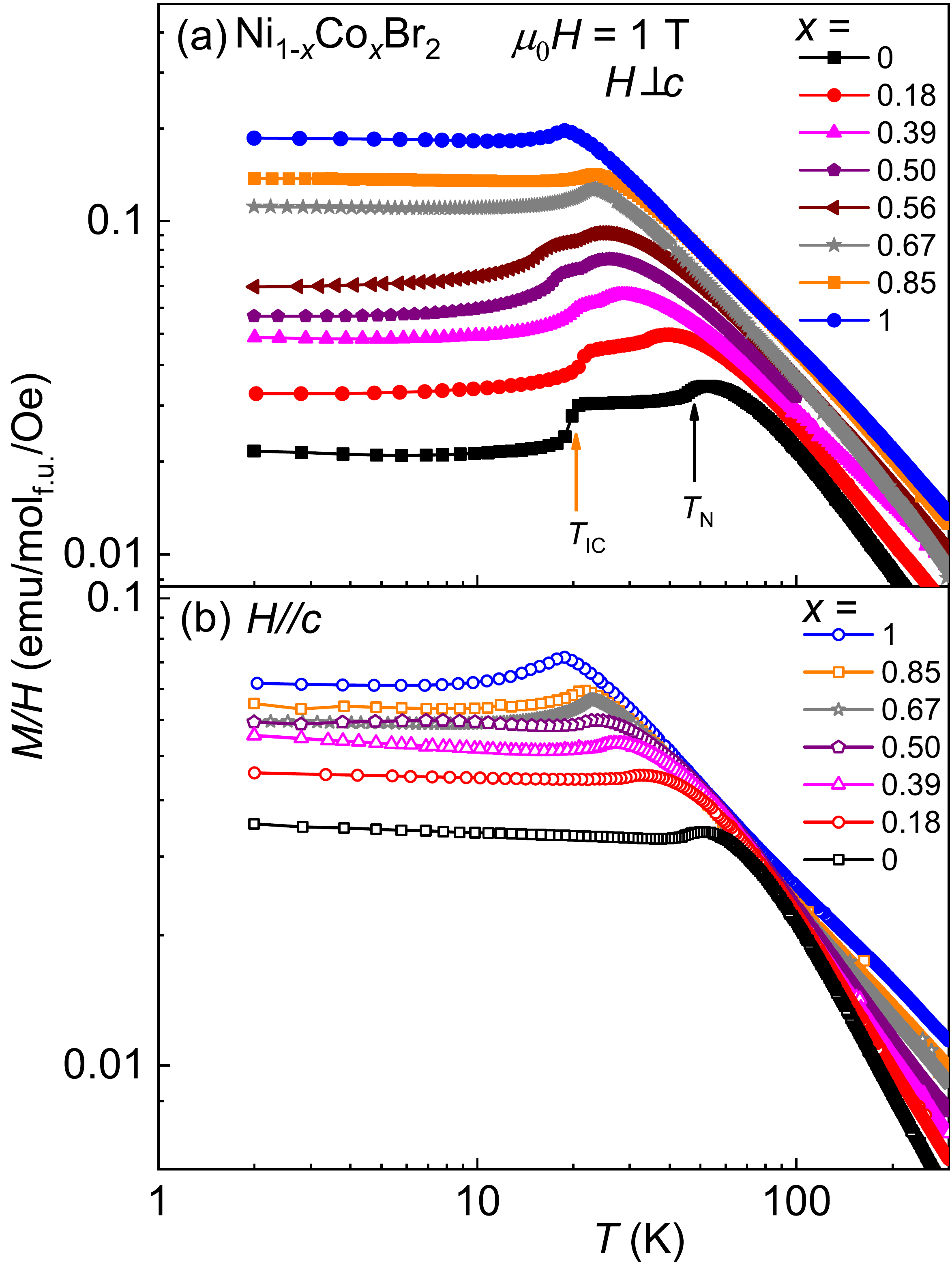}
\caption{\label{Fig2b} Temperature-dependent magnetization of Ni$_{1-x}$Co$_x$Br$_2$ collected upon cooling in an applied magnetic field $\mu_0 H$ = 1 T for (a) $H\perp c$ and (b) $H\parallel c$.}
\end{figure}

We have so far examined the temperature-dependent magnetization for NiBr$_2$ and CoBr$_2$. We now examine the impact of cobalt content on the magnetic properties. As shown in Fig.\ref{Fig2b}(a,b), the commensurate phase transition at $T_{\rm N}$ of Ni$_{1-x}$Co$_x$Br$_2$ is continually suppressed up to $x$ = 0.56, while a partial Ni substitution leads to an enhancement of $T_{\rm N}$ for CoBr$_2$. For $H\perp c$, the incommensurate phase transition ($T_{\rm IC}$ = 20(1) K) of NiBr$_2$ is first slightly enhanced with Co substitution and is last detected for $x$ = 0.56 (see in Fig. \ref{Fig2b}(a)).

\begin{figure}[htbp]
	\includegraphics[clip,width=\columnwidth]{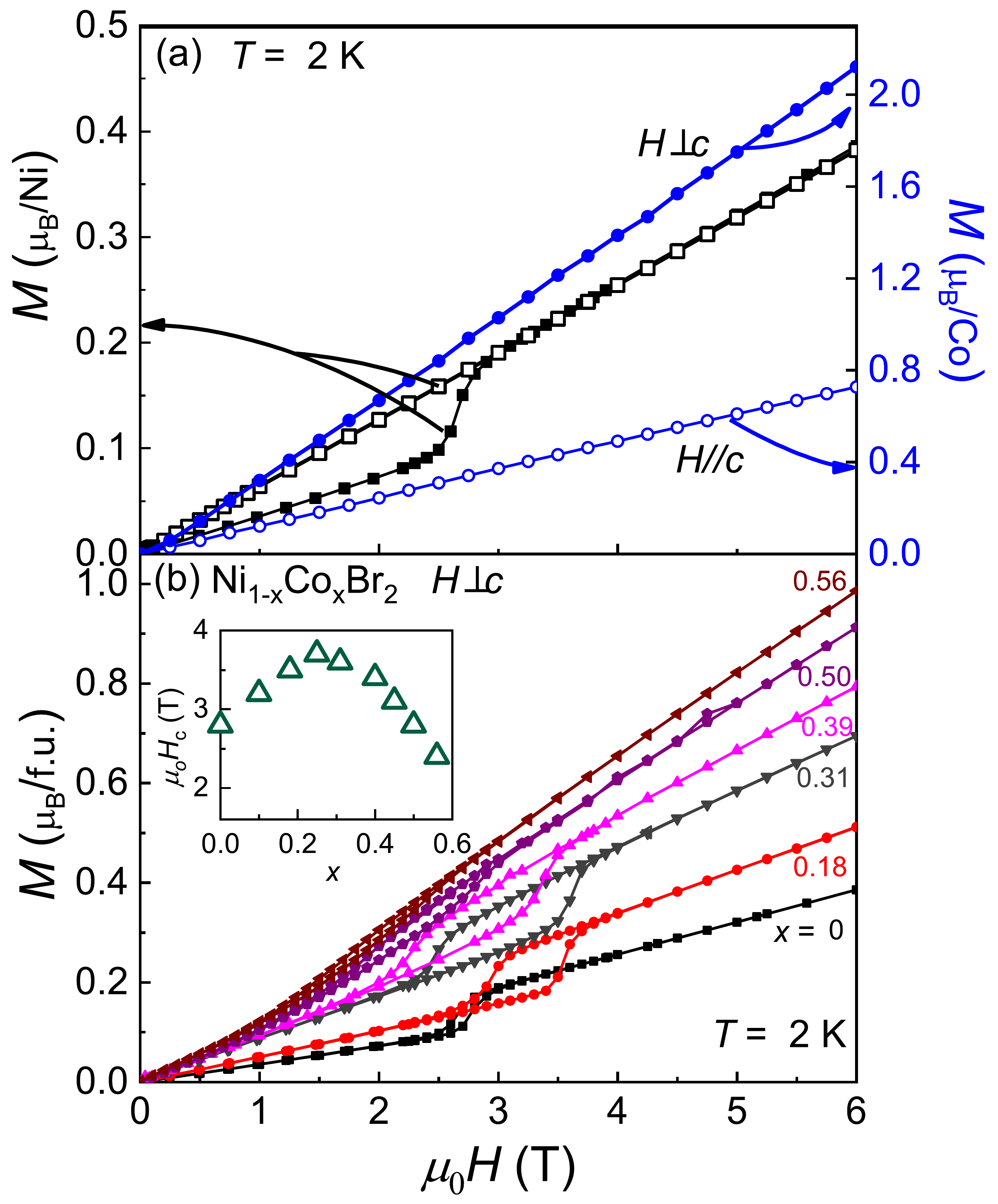}
\caption{\label{Fig3} Isothermal magnetization $M(H)$ at $T$ = 2 K. (a) $M(H)$ of NiBr$_2$ (left axis) and CoBr$_2$ (right axis) for $H\perp c$ (closed symbols) and $H\parallel c$ (open symbols). (b) $M(H)$ for selected compositions of Ni$_{1-x}$Co$_x$Br$_2$ with $H\perp c$. Inset: the critical magnetic field as a function of $x$ determined from a peak in d$M$/d$H$ for increasing applied magnetic field.}
\end{figure}

The isothermal magnetization $M(H)$ at $T$ = 2 K is shown in Fig. \ref{Fig3}. For a given magnetic field, $M$ of NiBr$_2$ (left axis in Fig. \ref{Fig3}(a)) is smaller than $M$ of CoBr$_2$ (right axis in Fig. \ref{Fig3}(a)). The induced magnetic moments are not saturated for $\mu_0 H$ = 6 T, and the maximum moment $M(\mu_0 H$ = 6 T, 2 K) increases continually from $x$ = 0 to 1. For NiBr$_2$, a magnetic anomaly around $\mu_0 H_c$ = 2.7 T is observed for $H\perp c$ (closed symbols), but it is absent for $H\parallel c$ (open symbols). The anomaly is related to a transition from the helimagnetic structure to the commensurate antiferromagnetic structure\cite{Day1981Moore, Tokunaga2011}. The observed magnetic hysteresis on increasing and decreasing magnetic fields (see in Fig. \ref{Fig3}(b)) is consistent with previous reports\cite{Adam1981,Tokunaga2011}, and suggests a first-order transition. Recently, Tokunaga et al. investigated multiferroic properties in NiBr$_2$ and reported a spontaneous polarization below $T_{\rm IC}$\cite{Tokunaga2011}. The ferroelectric transition disappeared above $H_c$, and the hysteresis may be related to magnetostructural domain movement.

\begin{figure}[htbp]
	\includegraphics[clip,width=0.95\columnwidth]{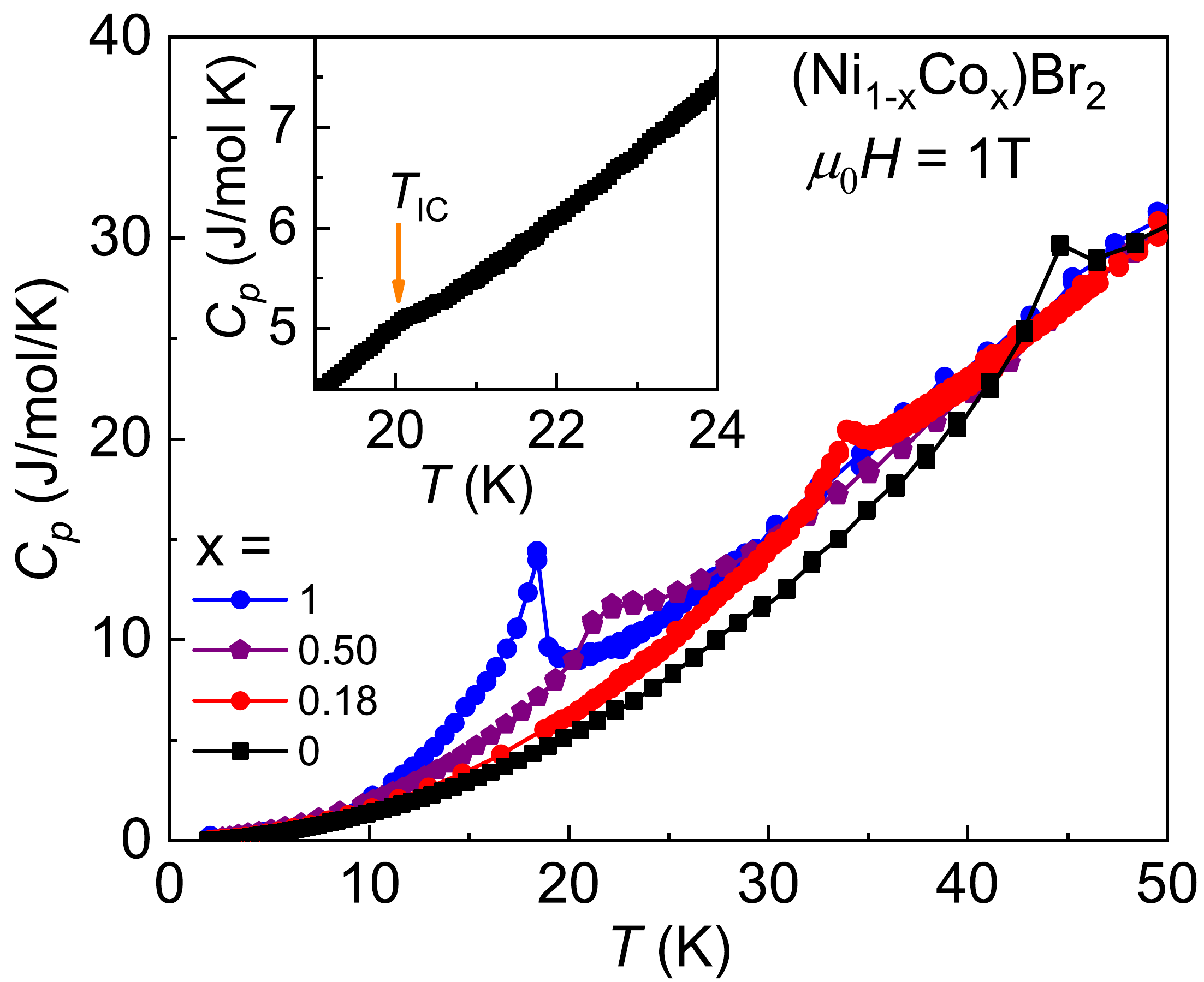}
\caption{\label{Cp} Specific heat $C_p(T)$ for Ni$_{1-x}$Co$_x$Br$_2$ for $\mu_0 H$ = 1 T applied within the $ab$ plane. Inset: a small anomaly observed at $T_{\rm IC}$.}
\end{figure}

Specific heat $C_p(T)$ of Ni$_{1-x}$Co$_x$Br$_2$ at  $\mu_0 H$ = 1 T are shown in Fig. \ref{Cp}. Strong anomalies are observed at $T$ = 44.6(2) K and  $T$ = 18.4(2) K for NiBr$_2$ and CoBr$_2$, respectively, which are consistent with $T_{\rm N}$ observed from temperature-dependent $M/H$. The commensurate-incommensurate phase transition observed in the magnetization for NiBr$_2$ is also observed in $C_p(T)$ as a weak anomaly around $T \approx$ 20 K (see in the inset of Fig. \ref{Cp}). We note that similar behavior was observed at $T_{\rm N}$ for specific heat in $H$ = 0, though the transition at $T_{\rm IC}$ was only observed in $C_p(T)$ with an applied magnetic field $H\perp c$ and thus results for $\mu_0H$ = 1 T are shown. $C_p(T)$ measurements were used to verify $T_{\rm N}$ of a few Ni$_{1-x}$Co$_x$Br$_2$ compositions.

\begin{figure}[t!]
	\includegraphics[clip,width=\columnwidth]{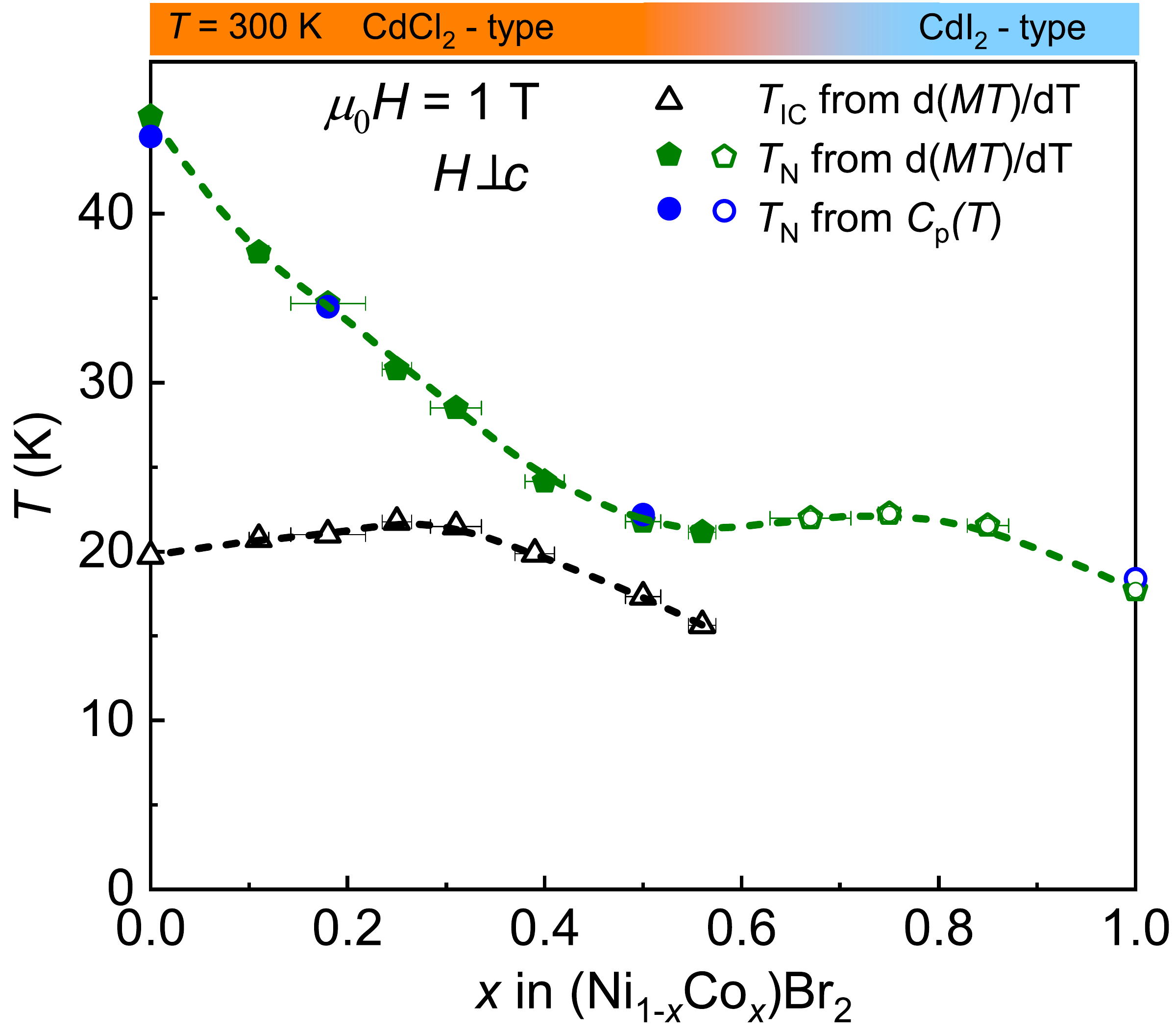}
\caption{\label{Phase} Temperature-composition phase diagram of magnetic transitions in Ni$_{1-x}$Co$_x$Br$_2$ for $H\perp c$ at $\mu_0 H$ = 1 T. The dashed lines guide the eye. The colored bar at the top of the graph represents the structural-phase diagram established from room-temperature x-ray diffraction data.}
\end{figure}

The magnetic transitions of the Ni$_{1-x}$Co$_x$Br$_2$ series derived from peaks in $C_p(T)$ and d($MT$)/d$T$\cite{Fisher1962} are summarized in the temperature-composition phase diagram shown in Fig. \ref{Phase}. Both magnetic transitions of NiBr$_2$ are affected by Co substitution. The commensurate phase transition is continually suppressed from $T_{\rm N} \approx$ 46 K for $x$ = 0 to $T_{\rm N} \approx$ 21 K for $x$ = 0.56. A slight enhancement of $T_{\rm N}$ for CoBr$_2$ is observed with a partial Ni substitution, and a local maximum of $T_{\rm N}$ occurs near $x$ = 0.75. Interestingly, $T_{\rm IC}$ shows a slight enhancement for small $x$, reaching a maximum at $T_{\rm IC}$ = 22(1) K for $x$ = 0.26 and then finally $T_{\rm IC}$ is suppressed beyond detection for $x >$ 0.56 at $\mu_0 H$ = 1 T. The evolution of $H_c$ as a function of cobalt content in $M(H)$ measurements follows a similar trend as $T_{\rm IC}$ in temperature-dependent $M/H$, indicating a strong correlation between $H_c$ and $T_{\rm IC}$. 

The presence of $T_{\rm IC}$ up to at least $x$ = 0.56 is robust compared to the behavior observed for other transition metal substitutions in NiBr$_2$. In contrast, the incommensurate magnetic structure of NiBr$_2$ disappears by $x$ = 0.112 in Ni$_{1-x}$Fe$_x$Br$_2$\cite{Moore1985}, although the CdCl$_2$ structure type is retained up to $x$ = 0.41. Interestingly, Ni$_{1-x}$Fe$_x$Br$_2$ with the helimagnetic structure and a finite easy-axis anisotropy is predicted to host magnetic field induced skyrmions\cite{Hayami2016}. Moreover, a neutron scattering study on Ni$_{0.92}$Zn$_{0.08}$Br$_2$ observed a ring of magnetic scattering in the $a^*b^*$ plane around (00$\frac{3}{2}$) and (10$\frac{\bar{1}}{2}$) resulting from a disordered propagation vector of the helimagnetic structure\cite{Day1981Moore}. It has been proposed that the ground state associated with this magnetic scattering is an impurity-driven vortex lattice phase\cite{Lin2016, Batista2016}. These experimental and theoretical results highlight the relevance of using chemical substitution in NiBr$_2$ to tune the exotic ground states\cite{Lin2016, Hayami2016}. As such, Ni$_{1-x}$Co$_x$Br$_2$ with the CdCl$_2$ structure type is a strong candidate to inspect for nontrivial spin textures like the magnetic vortex lattice phase or other multiple-$q$ spin textures. Examining the nature of the magnetic ground state of Ni$_{1-x}$Co$_x$Br$_2$ and the correlation between $T_{\rm IC}$ and $H_c$ requires additional characterization like neutron scattering, which is underway.

\section{Conclusions}

We synthesized single crystals of Ni$_{1-x}$Co$_x$Br$_2$ (0 $\leq~x~\leq$ 1) using vapor transport. A composition-induced transition from the CdCl$_2$ structure type of NiBr$_2$ to the  CdI$_2$ structure type of CoBr$_2$ occurs within 0.56 $<x<$ 0.76 at 300 K. The in-plane lattice parameter, layer spacing, and the average volume per atom continually increase with $x$. For Ni$_{1-x}$Co$_x$Br$_2$ with the CdCl$_2$ structure type, increasing the cobalt content leads to a suppression of the commensurate antiferromagnetic phase transition of NiBr$_2$. The transition temperature to the incommensurate phase is first slightly enhanced with cobalt substitution, then it is suppressed and last detected for $x$ = 0.56. The presence of $T_{\rm IC}$ up to at least $x$ = 0.56 indicates the delicate balance between ferromagnetic and antiferromagnetic exchange constants is perhaps still satisfied. Given that Ni$_{1-x}$Co$_x$Br$_2$ with the CdCl$_2$ structure type has both $T_{\rm N}$ and $T_{\rm IC}$ transitions in common with NiBr$_2$, which hosts the helimagnetic spin structure, these compositions appear as promising candidates in the ongoing search for topologically-nontrivial spin textures.

\section{Acknowledgements}

 This research was supported by the U.S. Department of Energy, Office of Science, Basic Energy Sciences, Materials Science and Engineering Division. We thank D. Parker and C. D. Batista for helpful discussions.

 \bibliography{Ni1-xCoxBr2}

\end{document}